\documentclass[
 reprint,
 superscriptaddress,
 bibnotes,
 amsmath,
 amssymb,
 aps,
 prb,
 citeautoscript,
 floatfix,
]{revtex4-2}

\usepackage{graphicx}
\usepackage{dcolumn}
\usepackage{bm}
\usepackage[dvipsnames]{xcolor}
\usepackage[colorlinks=true, citecolor={blue!80!black}, urlcolor={blue!50!black}, linkcolor = {blue!80!black}]{hyperref}
\usepackage{lipsum}  
\usepackage{upgreek}
\usepackage{siunitx}
\usepackage{mathtools}
\usepackage{verbatim}
\usepackage{xfrac}

\usepackage{nameref,zref-xr}
\zxrsetup{toltxlabel}

\newcommand{\abs}[1]{\left|#1\right|}
\newcommand{\ket}[1]{\left|#1\right\rangle}
\newcommand{\bra}[1]{\left\langle#1\right|}

\begin{document}
% \myexternaldocument{supplement}

\preprint{APS/123-QED}

\title{Flux-pulse-assisted Readout of a Fluxonium Qubit}

\author{Taryn V. Stefanski}
% \email[Correspondence email address:]{t.v.stefanski@tudelft.nl}
\affiliation{Quantum Engineering Centre for Doctoral Training, H. H. Wills Physics Laboratory and Department of Electrical and Electronic Engineering, University of Bristol, BS8 1FD, Bristol, UK}
\affiliation{QuTech and Kavli Institute of Nanoscience, Delft University of Technology, 2628 CJ, Delft, The Netherlands}

\author{Christian Kraglund Andersen}
\affiliation{QuTech and Kavli Institute of Nanoscience, Delft University of Technology, 2628 CJ, Delft, The Netherlands}

\date{16 April, 2024}

\begin{abstract}
Much attention has focused on the transmon architecture for large-scale superconducting quantum devices, however, the fluxonium qubit has emerged as a possible successor. With a shunting inductor in parallel to a Josephson junction, the fluxonium offers larger anharmonicity and stronger protection against dielectric loss, leading to higher coherence times as compared to conventional transmon qubits. The interplay between the inductive and Josephson energy potentials of the fluxonium qubit leads to a rich dispersive shift landscape when tuning the external flux. Here we propose to exploit the features in the dispersive shift to improve qubit readout. Specifically, we report on theoretical simulations showing improved readout times and error rates by performing the readout at a flux bias point with large dispersive shift. We expand the scheme to include different error channels, and show that with an integration time of 155 ns, flux-pulse-assisted readout offers about 5 times improvement in the signal to noise ratio.
Moreover, we show that the performance improvement persists in the presence of finite measurement efficiency combined with quasi-static flux noise, and also when considering the increased Purcell rate at the flux-pulse-assisted readout point. We suggest a set of reasonable energy parameters for the fluxonium architecture that will allow for the implementation of our proposed flux-pulse-assisted readout scheme.
\end{abstract}

\maketitle
\section{Introduction}
Two important prerequisites for meaningful quantum information processing are the ability to apply fast, accurate qubit gates and subsequently perform readout of the quantum state with high fidelity. High fidelity single shot measurement of superconducting qubits can be achieved through the exploitation of circuit quantum electrodynamics (cQED) \cite{blais2004cavity, schoelkopf2008wiring, wallraff2004strong}. By coupling a qubit to a far detuned resonator, where the qubit-resonator detuning is much larger than the coupling strength and resonator linewidth, we reach the dispersive limit. In this regime, we can perform a quantum non-demolition (QND) measurement through probing the readout resonator with coherent microwave pulses and measuring the transmitted and reflected signals that are correlated with the qubit state \cite{braginsky1980quantum, blais2004cavity, wallraff2004strong}. This method of measurement was initially implemented with a Cooper pair box and later in a transmon architecture, and can be further extended to other, novel superconducting device architectures such as the fluxonium \cite{wallraff2004strong, blais2004cavity, koch2007charge}. The fluxonium qubit, initially introduced by Ref.~\cite{manucharyan2009fluxonium}, is composed of a Josephson junction shunted by a capacitive and an inductive element. The qubit can be biased via an external magnetic flux applied through the loop made up of the non-linear Josephson junction and the inductor. Ideally, fluxonium qubits have maximal coherence and relaxation times when operated with an externally applied flux bias of $\Phi_{ext}\slash\Phi_{0} = 0.5$, also known as the ``sweet-spot,'' where $\Phi_{0} = h\slash(2e)$ is the magnetic flux quantum. At this point, the qubit is protected to first-order against flux noise and charge noise is exponentially suppressed \cite{manucharyan2009fluxonium, gyenis2021moving, lin2018demonstration, nguyen2022scalable, koch2009charging}.

As the interest in fluxonium qubits as a promising contender for quantum computation continues to grow due to their large anharmonicity and long coherence times, it is increasingly important to optimize all aspects of the fluxonium qubit, such as minimizing readout time while increasing the fidelity \cite{bao2022fluxonium, nguyen2022scalable}. At the half flux quantum point, it is desirable to have a small dispersive shift to mitigate dephasing due to residual photons in the readout resonator \cite{wang2019cavity}. However, this in turn makes fast readout challenging because longer integration times are required to differentiate the qubit states with high accuracy. To combat these opposing requirements, we propose the idea of using a rapid flux pulse to tune the qubit to a flux bias point where there is a large dispersive shift in order to perform the readout. Recent work with transmon qubits has demonstrated state-of-the-art measurement fidelity resulting, in part, from dynamic tuning of the qubit frequency in order to reduce the qubit-resonator detuning. This dynamical flux tuning subsequently increases the dispersive interaction strength \cite{swiadek2023enhancing}. This effect had been experimentally shown previously with transmon architectures in Refs.~\cite{sank2016measurement, krinner2022realizing}, and we now aim to expand the principle to fluxonium qubits. We also show this proposal to be robust, not only to variations in energy parameters due to fabrication imperfections, but also in the presence of limited measurement efficiency, quasi-static flux noise, and spontaneous emission into the resonator due to the Purcell effect. Ideally, the readout scheme should not be more sensitive to flux noise than other operations, so to put the simulations of quasi-static flux noise in context, we simulate optimized single qubit gates in the presence of this noise, as well \cite{bao2022fluxonium, moskalenko2022high, somoroff2021millisecond, nguyen2022scalable}.

In this work, we first present results of simulations performed to identify an appropriate energy parameter regime of the fluxonium qubit 
as a use case for implementation of fast readout via examination of the dispersive shift landscape. We proceed to explore how the signal to noise ratio and measurement fidelity are improved as a result of using a high dispersive shift during readout. Additionally, we consider limitations imposed by imperfect measurement efficiency, flux pulse duration, and Purcell decay. Finally, using the same energy parameters, we compare the effect of quasi-static flux noise on the error rates of both readout and single qubit gates.

\section{Fluxonium Qubit}
\begin{figure}
    \centering
    \includegraphics[scale=1.0]{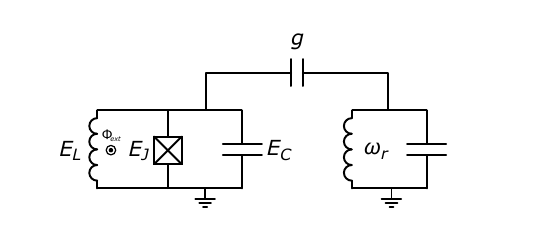}
    \caption{Effective circuit model of a fluxonium qubit capacitively coupled to a resonator, where $E_{J}$, $E_{C}$, and $E_{L}$ are the Josephson, capacitive, and inductive energies, respectively, $\Phi_{ext}$ is the flux threading the depicted loop, $g$ is the coupling strength between the qubit and readout resonator, and $\omega_{r}$ is the frequency of the readout resonator.}
    \label{fig:circuit}
\end{figure}

An effective lumped element circuit diagram of a fluxonium qubit coupled to a readout resonator is depicted in Fig.~\ref{fig:circuit}. The Hamiltonian used to describe the fluxonium qubit is
\begin{equation}
    \hat{H}_{\mathrm{fluxonium}} = 4E_{C}\hat{n}^2 + \frac{1}{2}E_{L}\hat{\phi}^2 - 
    E_{J}\mathrm{cos}(\hat{\phi}-\phi_{ext}),
    \label{Hflux}
\end{equation}

\noindent where $E_{C}$ is the charging energy of the fluxonium island, $E_{L}$ is the inductive energy of the inductor, $E_{J}$ is the Josephson energy, $\phi_{ext} = 2\pi(\Phi_{ext}\slash\Phi_{0})$, and $\Phi_{ext}$ is the physical flux threading the loop shared by the Josephson junction and the inductor. It is convenient to simulate the fluxonium Hamiltonian in the harmonic oscillator basis where the charge operator is defined as $\hat{n} = \frac{-i}{\sqrt{2}\phi_{0}}(\hat{c}-\hat{c}^{\dagger}$) and the flux operator takes the form $\hat{\phi} = \frac{\phi_{0}}{\sqrt{2}}(\hat{c}+\hat{c}^{\dagger})$, where $\phi_{0} = (8E_{C} \slash E_{L})^{1/4}$, and $\hat{c}$ and $\hat{c}^{\dagger}$ are the annihilation and creation operators, respectively, of the qubit basis states \cite{zhu2013circuit}.

The resonator and coupling between the resonator and qubit, where the coupling term assumes the rotating wave approximation, are modeled by
\begin{equation}
    \hat{H}_\mathrm{{resonator}} = \hbar\omega_{r}(\hat{a}^{\dagger}\hat{a}+\frac{1}{2}),
    \label{Hres}
\end{equation}
\vspace{-0.5cm}
\begin{equation}
    \hat{H}_\mathrm{{coupling}} = \hbar g(\hat{a}\hat{c}^{\dagger} + \hat{a}^{\dagger}\hat{c}).
    \label{Hcoup}
\end{equation}

\noindent In these equations, $\omega_{r} = 1 \slash \sqrt{LC}$ is the resonant frequency of the resonator, $g$ is the coupling strength mediated by the coupling capacitor, and $\hat{a}^{\dagger}$ and $\hat{a}$ are the creation and annihilation operators of the resonator. Summing the three terms expressed in Eq.~(\ref{Hflux}-\ref{Hcoup}) yields the total system Hamiltonian used in subsequent sections to obtain the numerical results.

\section{Dispersive Shift Landscape}
It is possible to simplify the total Hamiltonian by assuming the fluxonium qubit acts as a two level system, and thus use the Jaynes-Cummings model under the rotating wave approximation to describe the system in Fig.~\ref{fig:circuit}. Operating in the dispersive regime where the detuning between the qubit and resonator is much larger than the coupling strength ($\abs{\Delta} = \abs{\omega_{r}-\omega_{q}} \gg g$),  the Hamiltonian can be expanded to second order in $g\slash\Delta$ using perturbation theory \cite{schuster_2007}
\begin{equation}
    \hat{H}_\mathrm{{JC}} = \hbar\left(\omega_{r} + \frac{g^{2}}{\Delta}\hat{\sigma}_{z}\right)\hat{a}^{\dagger}\hat{a} + \frac{\hbar}{2}\left(\omega_{q} + \frac{g^{2}}{\Delta}\right)\hat{\sigma}_{z}.
    \label{Hjc}
\end{equation}

\noindent In this equation, $\omega_{q}$ represents the qubit frequency and $\hat{\sigma}_{z}$ is the Pauli-Z operator. 

From the first term in Eq.~\eqref{Hjc}, it becomes evident that the resonator frequency is dependent on the state of the qubit. The dispersive shift refers to the amount by which the qubit state dresses the bare resonator frequency and is denoted by $\chi = g^{2}\slash\Delta$. However, this approximate Hamiltonian neglects levels outside of the computational subspace which play an important role in the qubit-resonator interaction for fluxonium qubits \cite{zhu2013circuit}. Therefore, we instead numerically diagonalize the Hamiltonian for the qubit-resonator coupled system obtained from summing Eq.~(\ref{Hflux}-\ref{Hcoup}) in the harmonic oscillator basis and calculate the dispersive shift. We use the basis notation $\ket{\mathrm{qubit,resonator}}$, which identifies the energy of the dressed state closest to that of the corresponding bare state in order to obtain the dispersive shift, $\chi$, via
\begin{equation}
    2\chi = (\omega_{\ket{1,1}}-\omega_{\ket{1,0}}) - (\omega_{\ket{0,1}}-\omega_{\ket{0,0}}).
    \label{disshift}
\end{equation}
All simulations were performed using the NumPy and QuTiP Python libraries \cite{data, johansson2012qutip, johansson2013qutip2, harris2020array}. Calculations of relevant matrix elements, as will be discussed below, were done with the scQubits Python library \cite{groszkowski2021scqubits, chitta2022computer}.
\begin{figure*}
\includegraphics[scale=1]{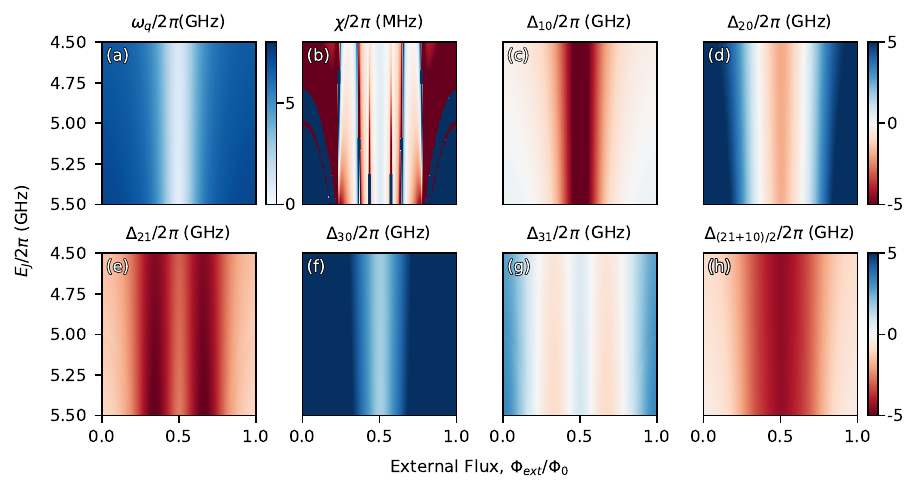}
\caption{(a) Qubit transition frequency, (b) dispersive shift, and (c-h) detuning  of various fluxonium transitions from the bare resonator frequency ($\Delta_{ij} = \omega_{ij}-\omega_{r}$) as a function of $E_{J}\slash2\pi$ (4.5-5.5 GHz) and external flux, where $E_{C}\slash2\pi = 1.25$ GHz, $E_{L}\slash2\pi = 1.5$ GHz, $\omega_{r}\slash2\pi = 7$ GHz, and  $g\slash2\pi = 50$ MHz. The maximum and minumum of (b) are $\pm$5 MHz and for (c-h) they are $\pm$5 GHz to provide visual clarity. Values outside of this range take on the color corresponding to the maximum or minimum value, depending on the sign. Panels (b-h) share the same color bar, which is depicted to the right of panels d and h.}
\label{fig:maps}
\end{figure*}

We aim to satisfy a moderately small dispersive shift on the order of $\chi\slash2\pi = 0.5-1$ MHz at $\Phi_{ext}\slash\Phi_{0} = 0.5$, as this gives a slow dephasing rate between approximately 4.5-16 kHz ($T_{2} \approx$ 60-220 $\mu$s in the limit as $T_{1} \rightarrow \infty$) with a conservative choice of effective resonator temperature of 60 mK and a resonator linewidth of $\kappa = 5$ MHz \cite{wang2019cavity}. At an optimal readout point, we desire a large dispersive shift on the order of $\chi\slash2\pi = 5-10$ MHz. Sub 100 ns readout has been demonstrated with dispersive shift values in this range for transmon qubits \cite{walter2017rapid}. Another criterion to satisfy is having minimal avoided crossings in the dispersive shift between $\Phi_{ext}\slash\Phi_{0} = 0.5$ and the chosen readout point. This requirement is to avoid coherently swapping excitations between the resonator and qubit while flux tuning to the readout point.

In Fig.~\ref{fig:maps}(a) and (b), we numerically solve for the fluxonium qubit transition frequency, $\omega_{q}$ = $
\omega_{\ket{1,0}}-\omega_{\ket{0,0}}$, and for the dispersive shift, $\chi$, respectively. While the behavior of $\omega_{q}$ is rather smooth, we notice many rapidly changing features in $\chi$. Since the fluxonium qubit is not restricted by stringent selection rules away from flux biases equalling half-integer and integer flux quanta, higher order transitions can interact with the readout resonator, giving rise to a large dispersive shift (see Fig.~\ref{fig:maps}(b-h)). In Fig.~\ref{fig:maps}(c-h), we display the detunings between various dressed, higher order transitions and the bare readout resonator frequency, where the detuning is defined as
\begin{equation}
    \Delta_{ij} = \omega_{ij} - \omega_{r},
    \label{detuning}
\end{equation}
\noindent with
\begin{equation*}
    \omega_{ij} = \omega_{\ket{i,0}}-\omega_{\ket{j,0}}.
    \label{subcripts}
\end{equation*}
\noindent Ref.~\cite{zhu2013circuit} elucidated the fact that each higher order transition contributes to the effective dispersive shift. The dispersive shift caused by a given qubit transition $\ket{i} \leftrightarrow \ket{j}$ is proportional to $\abs{\bra{i}\hat{n}\ket{j}}^{2}\slash\Delta_{ij}$. Thus, as $\Delta_{ij} \rightarrow 0$, we expect an increase in the magnitude of the dispersive shift. At sufficiently small detunings, the interaction is no longer considered dispersive and, thus, we want to avoid these regions as the dynamics will be dominated by a coherent swap interaction between the fluxonium and the resonator. By matching the regions of $\Delta_{ij} = 0$ in Fig.~\ref{fig:maps}(c-h) with regions of diverging dispersive shift, $\chi$, in Fig.~\ref{fig:maps}(b), we can identify which specific transition causes each divergence in the dispersive shift. For instance, the two dark, narrow lines in Fig.~\ref{fig:maps}(b) closest to the half flux quantum point are caused by the $\ket{3} \leftrightarrow \ket{1}$ transition being on resonance with the readout resonator. 

\section{Readout}
Fluxonium qubits with frequencies around 1 GHz or lower at the sweet-spot have shown consistently high $T_{1}$ times exceeding 100 $\mu$s \cite{nguyen2019high}. To satisfy this requirement on the qubit frequency, as well as the requirements for the dispersive shift discussed in the previous section, we pick appropriate energy parameters for the fluxonium based on the results in Fig.~\ref{fig:maps}. We propose to use a fluxonium qubit with the parameters $E_{J}\slash2\pi = 4.75$ GHz, $E_{C}\slash2\pi = 1.25$ GHz, and $E_{L}\slash2\pi = 1.5$ GHz for implementing a flux pulse during readout to improve fidelity and speed. These values were chosen to keep both the qubit frequency and dispersive shift low at the sweet-spot, as altering one of the energy parameters results in a decrease in one value, but increase in the other. Additionally, targeting these values ensures that in the case of deviations due to fabrication imperfections, the proposed readout scheme can still be implemented (see Appendix A). We note that the fluxonium qubit is symmetric about the sweet-spot, so we focus our discussion on values of $\Phi_{ext}\slash\Phi_{0} \ge 0.5$ for the rest of this section.

\begin{figure}
    \centering
    \includegraphics[scale=1]{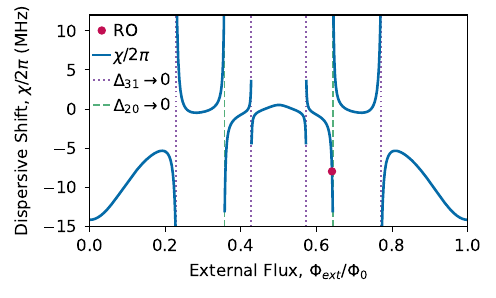}
    \caption{Dispersive shift as a function of external flux where $E_{J}\slash2\pi = 4.75$ GHz, $E_{C}\slash2\pi = 1.25$ GHz, and $E_{L}\slash2\pi = 1.5$ GHz. The targeted readout (RO) point is marked with a red circle, the blue curve shows $\chi\slash2\pi$, and the dashed and dotted vertical asymptotes denote where a certain qubit transition, $\ket{i} \leftrightarrow \ket{j}$, and the resonator are on resonance, denoted by $\Delta_{ij}\rightarrow0$ in the legend.}
    \label{fig:1d chi}
\end{figure}

In order to simulate our system with the specified parameters, we use a readout resonator frequency of $\omega_{r}\slash2\pi = 7$ GHz and a coupling strength of $g\slash2\pi = 50$ MHz. At the half flux quantum point, $\chi\slash2\pi \approx 0.53$ MHz and $\omega_{q}\slash2\pi \approx 1.05$ GHz. The proposed readout point is denoted with a red circle in Fig.~\ref{fig:1d chi}, which sits at a flux bias of $\Phi_{ext}\slash\Phi_{0} \approx 0.64$, where $\chi\slash2\pi \approx -8$ MHz, $\omega_{q}\slash2\pi \approx 4.6$ GHz, and $\Delta_{10}\slash2\pi = -2.4$ GHz. The large dispersive shift arises due to $\Delta_{20}$ approaching zero. 

At the specific value we propose, we extract $\Delta_{20}\slash2\pi = -67$ MHz and $g_{20}\slash2\pi$ = $(g\slash2\pi)\abs{\bra{2}\hat{n}\ket{0}}$ = 18.32 MHz. This yields a contribution to $\chi$ of $g_{20}^{2}\slash\Delta_{20} = 2\pi\times-5$ MHz and places the $\ket{2} \leftrightarrow \ket{0}$ qubit subspace on the edge of the dispersive limit. Importantly, the computational subspace of the qubit remains well into the dispersive regime at this point with $g_{10}\slash2\pi$ = $(g\slash2\pi)\abs{\bra{1}\hat{n}\ket{0}}$ = 18 MHz so that $\Delta_{10} \gg g, g_{10}$. It should also be noted that while the dispersive shift more gradually approaches a large magnitude around integer multiples of flux quanta, we do not perform readout near these points as the large values are a direct result of small detuning between the $\ket{1}\leftrightarrow\ket{0}$ qubit transition and resonator frequency.

From Fig.~\ref{fig:1d chi}, it is evident that we must traverse a point at which the $\ket{3}\leftrightarrow\ket{1}$ qubit transition is resonant with the resonator around $\Phi_{ext}\slash\Phi_{0} \approx 0.575$. We numerically calculate $g_{31}\slash2\pi = 5.81$ MHz from the size of the anticrossing, which yields a swap time of roughly 43 ns. From the Jaynes-Cummings model, the swap time is defined as $\pi\slash2g_{ij}$, and is the time it takes for an excitation to be exchanged between the qubit subspace and resonator, given they are on resonance. The swap time sets an upper limit on the ramp time of the flux pulse. However, there is a trade-off between decreasing the ramp time to ensure the QNDness of the readout is maximized (see Appendix B), and having a long enough ramp time for the process to remain adiabatic. To be considered adiabatic, the ramp time must be on a time scale $\gg 2\pi\slash\omega_{q} \approx$ 1 ns \cite{Sakurai_Napolitano_2020}.

In order to achieve high measurement fidelity, we aim to perform a strong measurement in a short amount of time so that we avoid relaxation of the qubit of interest. We can quantify the strength of the measurement using the signal to noise ratio (SNR) as a metric. This value is the ratio between the separation of the mean of the $\ket{0}$ and $\ket{1}$ state projections in the (I,Q) plane, and the total standard deviation of the two signal distributions which typically take on a Gaussian profile \cite{clerk2010introduction, krantz2019quantum}. The separation between the distributions grows linearly in time, $t$, compared to the width of the distributions which scales as $\sqrt{t}$ \cite{clerk2010introduction}. We are in the strong measurement regime when the SNR is greater than unity and thereby, the two states can be distinguished because the distributions have separated after a certain amount of integration time.

We can use the Langevin equation in the context of input-output formalism in order to numerically simulate the output field of a resonator when probed with a coherent input field via a capacitively coupled drive line. This is expressed as 
\begin{equation}
    \dot{\hat{a}} = -i\chi\hat{\sigma}_{z}\hat{a} - \frac{1}{2}\kappa\hat{a} - \sqrt{\kappa}\hat{a}_{in},
    \label{langevin}
\end{equation}
\vspace{-0.5cm}
\begin{equation}
    \hat{a}_{out} = \hat{a}_{in} + \sqrt{\kappa}\hat{a},
    \label{langevinbc}
\end{equation}

\noindent where Eq.~\eqref{langevinbc} is the boundary condition for Eq.~\eqref{langevin}, $\hat{a}_{in}$ is the coherent input field such that $\langle \hat{a}_{in} \rangle$=$-\epsilon\slash\sqrt{\kappa}$, $\hat{a}$ is the intracavity field, $\hat{a}_{out}$ is the output field, and $\kappa$ is the photon decay rate into the drive line or the linewidth of the resonator \cite{gardiner2004quantum, didier2015fast, didier2015heisenberg}. Assuming the resonator is in a coherent state as we probe it with a classical signal, Eq.~\eqref{langevin} can alternatively be solved in terms of the coherent state amplitude, $\alpha$, from 
\begin{equation}
    \dot{\alpha} = -i\chi\left\langle\hat{\sigma}_{z}\right\rangle\alpha - \frac{1}{2}\kappa\alpha - \sqrt{\kappa}\alpha_{in},
    \label{langevinalpha}
\end{equation}

\noindent where $\alpha_{(in/out)}$ = $\left\langle\hat{a}_{(in/out)}\right\rangle$. Eq.~\eqref{langevinalpha} has the solutions
\begin{equation} \begin{split}
    &\alpha_{\pm} = \frac{\epsilon}{\kappa}\biggl[1 + 
e^{-i\phi_{qb}\left\langle\hat{\sigma}_{z}\right\rangle} \\
    &\times\left(1-2 \mathrm{cos}\left(\frac{\phi_{qb}}{2}\right)e^{\left(-i\chi t \left\langle\hat{\sigma}_{z}\right\rangle-\frac{\kappa t}{2}+i\frac{\phi_{qb}}{2}\left\langle\hat{\sigma}_{z}\right\rangle\right)}\right)\biggr].
    \label{alphasolution}
\end{split}\end{equation}

\noindent From Eq.~\eqref{alphasolution}, $\alpha_{+}$ ($\alpha_{-}$) is obtained when $\left\langle\hat{\sigma}_{z}\right\rangle$ = +1 (-1), corresponding to the qubit eigenstate $\ket{0}$ ($\ket{1}$). Additionally, $t$ is the integration time, $\epsilon$ is the driving amplitude of the input field, and $\phi_{qb}$ is the phase shift of the output field caused by the qubit, defined as
\begin{equation}
    \phi_{qb} = 2 \mathrm{arctan}\left(\frac{2\chi}{\kappa}\right).
\end{equation}

\noindent We can choose $\epsilon$ based on the targeted mean photon number by solving for $\overline{n}$ in the steady state in terms of the drive amplitude
\begin{equation}
    \epsilon = \sqrt{\overline{n}\left(\frac{\kappa^{2}}{4}+\chi^{2}\right)}.
    \label{res drive amp}
\end{equation}

\noindent Once we have found $\alpha_{\pm}$, we can find $\alpha_{out}$ via the equation
\begin{equation}
    \alpha_{out} = \alpha_{in}+\sqrt{\kappa}\alpha.
\end{equation}

\noindent We can describe the signal to noise ratio in terms of the measurement operator $\hat{M}$
\begin{equation}
    \hat{M}(\tau) = \sqrt{\kappa\eta}\int_0^\tau\hat{a}_{out}(t) dt,
\end{equation}

\noindent where we assume heterodyne detection and use $\eta$ to represent the measurement efficiency. We denote $\left\langle\hat{M}\right\rangle$ as $M_{S}$ and the noise operator as $\hat{M}_{N}$ = $\hat{M}$-$\left\langle\hat{M}\right\rangle$, following the convention in Ref.~\cite{didier2015heisenberg}. The equation describing the signal to noise ratio then reads as
\begin{equation}
    SNR = \frac{|M_{S,\eta,\ket{0}}-M_{S,\eta,\ket{1}}|}{\left[\left\langle\hat{M}_{N,\ket{0}}^{2}\right\rangle+\left\langle\hat{M}_{N,\ket{1}}^{2}\right\rangle\right]^{1/2}}.
\end{equation}

\noindent We can evaluate $\left\langle\hat{M}_{N}^{2}\right\rangle$ to $\kappa\tau\slash2$, where $\tau$ is the total integration time, since we are using a coherent input field.

\begin{figure}
    \centering
    \includegraphics[scale=1]{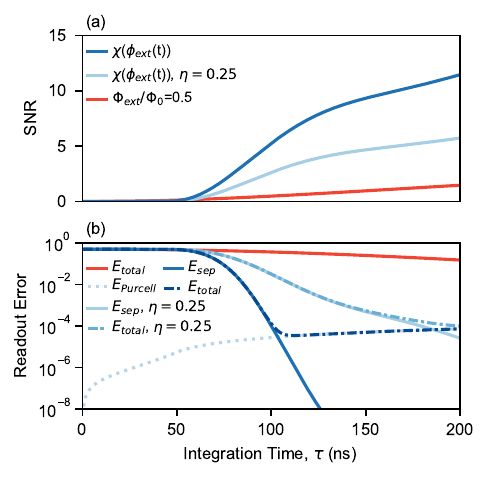}
    \caption{Comparison of readout at the half flux quantum point, where $\Phi_{ext}\slash\Phi_{0} = 0.5$ and $\chi\slash2\pi = 0.527$ MHz (red), versus readout incorporating a flux pulse with 50 ns rise time to $\Phi_{ext}\slash\Phi_{0} = 0.641$ with $\chi\slash2\pi = -7.95$ MHz for measurement efficiencies of 25$\%$ (light blue, solid) and 100$\%$ (dark blue, solid). (a) Signal to noise ratio versus integration time. (b) Readout error as a function of integration time. Solid curves show the readout error limited only by SNR ($E_{sep}$). The dotted, light blue line marks the error arising from Purcell decay in the flux-pulse-assisted readout case ($E_{Purcell}$). Dash-dotted lines show the total error rates ($E_{total}$) from summing the two contributions for $
    \eta$ = 25\% (light) and $\eta$ = 100\% (dark).}
    \label{fig:ro compare}
\end{figure}

Combining these formulas, we first calculate the SNR for conventional fluxonium readout where $\Phi_{ext}\slash\Phi_{0} = 0.5$ and remains static, see red lines in Fig.~\ref{fig:ro compare}. At the half flux quantum point, the dispersive shift is $\chi\slash2\pi = 0.527$ MHz and with a resonator linewidth of $\kappa\slash2\pi = 5$ MHz, perfect measurement efficiency, and $\Bar{n} = 10$, we obtain an SNR of 1 after about 155 ns. 

Following our proposal, we next discern the effect of performing readout during flux tuning of the qubit. We assume a flux pulse rise time of 50 ns and assume the flux pulse remains on for the remaining duration of the readout. We numerically solve the Langevin equation using an explicit time dependent dispersive shift $\chi(\phi_{ext},t) = \chi(\phi_{ext}(t))$, following the behavior shown in Fig.~\ref{fig:1d chi} and assuming the flux increases linearly from $\Phi_{ext}\slash\Phi_{0} = 0.5$ to the target flux of $\Phi_{ext}\slash\Phi_{0} = 0.641$ during the first 50 ns of the readout. During this time, we also quadratically increase the resonator drive amplidude, $\epsilon$, until we reach the value given by Eq.~\eqref{res drive amp}. Because we are reading out at a frequency equal to the bare resonator frequency, the gradual rise of the drive amplitude avoids strongly populating the readout resonator at the start of the readout where the dispersive shift is low.

The SNR slowly increases during the ramp time of the flux pulse and then increases more rapidly once the steady value of $\Phi_{ext}\slash\Phi_{0} = 0.641$ is reached, see blue lines in Fig.~\ref{fig:ro compare}(a). Assuming 100$\%$ measurement efficiency, the SNR is around 9.5 after 155 ns, giving almost an order of magnitude improvement in the SNR as compared to static readout at the sweet-spot. With a realistic experimental measurement efficiency of 25$\%$, after 155 ns the SNR is still about 5 times larger than the SNR obtained with perfect measurement efficiency when readout is performed at the sweet-spot. Thus, we can more rapidly distinguish the two qubit states with high certainty via the exploitation of a strong dispersive shift. We use a long flux pulse rise time in these simulations to numerically demonstrate proof of principle and show that we still see a substantial improvement in measurement time, and potential additional speedup could be achieved with flux pulse parameter optimization \cite{bryon2022experimental, chen2022fast}.

From the signal to noise ratio, we can also extract the SNR-limited, or separation, error from
\begin{equation}
    E_{sep} = \frac{1}{2}\mathrm{erfc}\left(\frac{\mathrm{SNR}}{2}\right),
\end{equation}
\noindent with
\begin{equation*}
    \mathrm{erfc}(x) = 1-\frac{2}{\sqrt{\pi}}\int_{0}^{x}e^{-t^2}dt,  
\end{equation*}

\noindent which is modeled by the solid curves in Fig.~\ref{fig:ro compare}(b).

As expected, the readout error trends towards zero for large times, although much more rapidly for the case where we flux tune to the suggested readout point. Using the proposed flux pulsing scheme, an error of $10^{-3}$ is reached in roughly 140 ns for the case of 25$\%$ measurement efficiency, compared to 570 ns if readout is performed at the half flux quantum point with perfect measurement efficiency. 

We also consider the limitation imposed by an increased Purcell decay rate at the readout point compared to that at sweet-spot. In the dispersive interaction picture, the Purcell rate of decay is modeled by \cite{houck2008controlling, sete2015quantum}
\begin{equation}
    \Gamma_{\mathrm{P}} = \left(\frac{g}{\Delta}\right)^{2}\kappa.
    \label{purcell}
\end{equation}
\noindent We focus our attention on the qubit's computational subspace and thus use $g_{10}$ and $\Delta_{10}$ in evaluating Eq.~\eqref{purcell}. In this context, $\Delta_{10}$ is the difference between the bare qubit and bare resonator frequencies. Because Purcell decay only leads to an incorrect measurement of the ground state if the decay occurs at a time $<\tau\slash$2, the contribution to the readout error can be expressed as \cite{sete2015quantum, khezri2015qubit}
\begin{equation}
    E_{Purcell} = \frac{1}{4}(1-e^{-\int_{0}^{\tau}\Gamma_{\mathrm{P}}(t) dt}).
    \label{Purcell}
\end{equation}

At the sweet-spot, we find a Purcell decay time of $T_{1,\mathrm{P}} \approx$ 11 ms, which means that the achievable readout error only becomes Purcell-limited for integration times $\gtrsim$ 800 ns. At the flux-pulse-assisted readout point, the qubit-resonator detuning is reduced, yielding a decreased Purcell decay time of $T_{1,\mathrm{P}} \approx$ 560 $\mu$s. Flux-pulse-assisted readout becomes Purcell-limited for integration times beyond about 150 ns(100 ns) with $\eta$ = 25\%(100\%). Despite the increased Purcell decay rate, flux-pulse-assisted readout outperforms readout at sweet-spot for integration times up to about 650 ns, and offers an advantage of at least 2 orders of magnitude for times between $\sim$150-400 ns. One could consider employing a Purcell filter within the circuit design if further suppression of Purcell decay is desired \cite{reed2010fast, sete2015quantum, heinsoo2018rapid}.

In this section, we focus our analysis on prominent error channels that are intrinsic to the readout design and parameters. It is important to note that in practice we will also be limited by the finite relaxation time of the qubit (see Appendix C) and potential measurement-induced state transitions, which were not accounted for in these simulations. The occurrence of measurement-induced state transitions are expected to degrade the QNDness of the readout and limit the measurement fidelity. However, simulating the precise nature and conditions of these transitions, as has been done extensively in the context of transmon qubits, goes beyond the scope of this present work \cite{sank2016measurement, khezri2023measurement, shillito2022dynamics, dumas2024unified, nesterov2024measurement}. Finally, due to the low frequency of the fluxonium at sweet-spot which leads to a large thermal population, any characterization of readout would be limited by the initialization fidelity. Various methods of initialization have been employed experimentally in fluxoniums with high fidelity \cite{bao2022fluxonium, ding2023high, gusenkova2021quantum, johnson2012heralded, moskalenko2022high, zhang2021universal}. Here, we assume that the qubit is perfectly initialized. 

\section{Quasi-static Flux Noise}
Although the fluxonium qubit benefits from long coherence times partially afforded by the first-order insensitivity to flux noise at the sweet-spot, our proposed protocol requires flux-pulsing to a bias point that has an increased susceptibility to flux noise. The prevalent quasi-static 1/$f$ noise, varying from one pulse sequence to the next, results in qubit dephasing \cite{bylander2011dynamical}. While additional dephasing during readout is irrelevant as a strong measurement inherently dephases the qubit, we need to verify that the flux-pulse-assisted measurement performance is not limited by potential flux noise.

\begin{figure}
    \centering
    \includegraphics[scale=1]{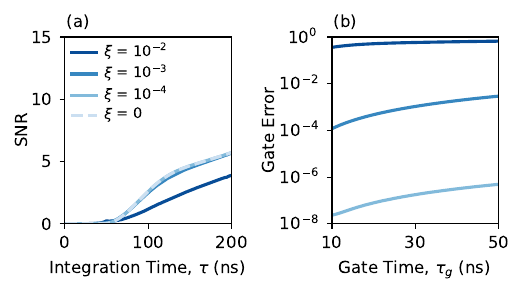}
    \caption{Comparison of (a) readout signal to noise ratio obtained using flux-pulse-assisted readout with a measurement efficiency of 25$\%$ and (b) single qubit gate fidelity of a Pauli-X gate with varying amounts of quasi-static flux noise added in the form of a constant offset, $\delta$, where $\delta$ = $\xi x\Phi_{0}$ and $x$ is a value sampled randomly from a normal Gaussian distribution.}
    \label{fig:flux noise}
\end{figure}

In this section, we simulate the presence of quasi-static flux noise by adding a constant flux bias offset of $\delta$ during flux-pulse-assisted readout, assuming 25$\%$ measurement efficiency. We average over 50 iterations of $\delta$ for 3 different scaling factors. In this context, $\delta$ = $\xi x\Phi_{0}$, where $\xi$ is a scaling factor of $10^{-2}, 10^{-3}, \mathrm{or}$ $10^{-4}$, and $x$ is a random number sampled from a Gaussian distribution with a mean of $\mu = 0$ and a variance of $\sigma^{2} = 1$, for each iteration. The literature commonly reports flux noise amplitudes on the order of 1-10 $\mu\Phi_{0}$ observed in superconducting qubit devices \cite{braumuller2020characterizing, kou2017fluxonium, gustavsson2011noise, yoshihara2006decoherence}. Thus, the amount of noise we add to our numerical simulations reflects flux noise significantly larger than that observed in state-of-the-art experiments. As displayed in Fig.~\ref{fig:flux noise}(a), for the extreme case of $\xi$ = $10^{-2}$, our SNR after 155 ns is still almost 3 times larger than the SNR achieved with a static readout at sweet-spot with perfect measurement efficiency (refer to red line of Fig.~\ref{fig:ro compare}(a)). This corresponds to a readout error of roughly 8$\%$. In the cases of $\xi$ = $10^{-3}$ or $10^{-4}$, there is no substantial deviation from the readout achievable in the presence of zero quasi-static flux noise. It should be noted that the small peak around 50 ns for $\xi$ = $10^{-2}$ is an artifact of averaging over iterations where the flux pulse overshoots the targeted readout point such that it is closer to the $\ket{2}\leftrightarrow\ket{0}$ avoided crossing, yielding an extremely large dispersive shift magnitude. The quasi-static flux noise offset does not improve the readout SNR for short integration times, generally.

To put these numbers into context, we also simulate the single qubit gate fidelity of a Pauli-X gate with a DRAG pulse performed at the sweet-spot in the presence of varying degrees of quasi-static flux noise in order to further characterize the achievable quality of qubit operations in the presence of this decoherence channel. Thus, our simulations serve to verify that our flux-pulse-assisted readout scheme is not the limiting factor for overall qubit performance. We add the flux noise offset in the same manner which was done for the readout simulations previously discussed. Prior to adding the noise, we first optimize the performance of the $X_{\pi}$ rotation. 

In order to model these single qubit gates, we add a drive term to the Hamiltonian of the fluxonium-resonator coupled system (Eq.~(\ref{Hflux}-\ref{Hcoup})). We assume the qubit is capacitively coupled to a microwave drive line that receives signals from room temperature electronics, of which experimental parameters including pulse shape, amplitude, duration, frequency, and phase can be altered.

The envelope function used to describe the shape of the in-phase pulse component is
\begin{equation}
    s(t) = \frac{1}{2}\left(1-\mathrm{cos}\left(\frac{2\pi t}{\tau_{g}}\right)\right).
\end{equation}

\noindent We opt for a sinusoidal envelope function as in Ref.~\cite{nguyen2022scalable} in order to avoid the truncation that is necessary with a traditional Gaussian envelope function and to reduce the bandwidth of the pulse. A derivative removal by adiabatic gate (DRAG) pulse is also applied in the out-of-phase component to help correct for leakage and pulse errors, as has been done before in transmon qubits and has been theoretically explored in scalable fluxonium architectures \cite{motzoi2009simple, krantz2019quantum, gambetta2011analytic, nguyen2022scalable}. The envelope function of this pulse takes the form
\begin{equation}
    s'(t) = \frac{\lambda}{\alpha}\frac{\partial s(t)}{\partial t},
\end{equation}

\noindent where $\lambda$ is a scaling parameter and $\alpha$ is the qubit's anharmonicity.

With this pulse shape defined, we can write the drive term of the Hamiltonian as 
\begin{equation}
    \hat{H}_{\mathrm{drive}} = \varepsilon_{d}\hat{n}[2s(t)\mathrm{sin}(\omega_{d}t)
    + s'(t)\mathrm{cos}(\omega_{d}t)],
\end{equation}

\noindent where $\varepsilon_{d}$ is the drive amplitude and $\omega_{d}$ is the frequency of the drive signal.

We first optimize over $\varepsilon_{d}$ and $\lambda$, assuming that $\omega_{d} = \omega_{q, \varphi_{ext} = \pi}$, for gate times ranging from 10-50 ns. We subsequently add the flux noise offset to the single qubit gate performance with optimized parameters. These results are presented in Fig.~\ref{fig:flux noise}(b). In the noisiest case simulated with $\xi$ = $10^{-2}$, our single qubit gate error has a high value of 36$\%$ with a gate time of 10 ns, which is an order of magnitude worse than the readout error induced by this level of flux noise. If we compare the single qubit gate error curves in the cases of $\xi$ = $10^{-3}$ and $10^{-4}$, we see that one order of magnitude increase in quasi-static flux noise degrades the gate error by 4 orders of magnitude. Thus, we observe that our flux-pulse-assisted readout scheme is more robust than optimized single qubit gates in the face of quasi-static 1/$f$ flux noise.

\section{Conclusion and Outlook}
In summary, we present a proposal for performing flux-pulse-assisted, fast, high fidelity readout of a fluxonium qubit with one set of optimized energy parameters. We suggest appropriate values of Josephson, capacitive, and inductive energies for a fluxonium qubit which will enable our readout proposal based on the simulated dispersive shift landscape. Future work may consider exploring the implementation of this scheme in more novel energy regimes, such as the heavy fluxonium \cite{zhang2021universal, earnest2018realization}.

We demonstrate the improvement in measurement speed and fidelity that can be obtained by exploiting a large dispersive shift when the qubit is detuned from the sweet-spot, despite an increased Purcell decay rate and susceptibility to flux noise at this point. In this work, we do not explicitly consider swap dynamics or changes in the dispersive shift due to AC Stark shifting during measurement, however, we expect this can be compensated for through flux pulse calibration. Pulse shaping techniques may also be utilized to further reduce the time required for the scheme. We present the readout SNR and single qubit gate fidelity achievable in the presence of varying degrees of quasi-static flux noise. Finally, we compare the contribution of various error channels discussed in the text in the context of both readout at sweet-spot and with flux-pulse-assisted readout. These quantities are listed in Table~\ref{tab:error summary} with the addition of error resulting from thermal excitation. Our simulated results show robust improvement in both readout speed and fidelity when the proposed protocol is implemented. 

\begin{table}[]
\caption{Summary of readout errors for readout at sweet-spot compared to the flux-pulse-assisted readout scheme after an integration time of 200 ns. The bottom row quantifies the level of non-QNDness expected to arise from traversing the $\ket{3}\rightarrow\ket{1}$ avoided crossing during the flux pulse ramp.}
\label{tab:error summary}
\renewcommand{\arraystretch}{1.25}
\setlength{\tabcolsep}{4.5pt}
\begin{tabular}{lcc}
\hline
\hline
Error & Sweet-spot & Flux-pulse-assisted \\ \hline
Purcell decay & 4.5$\times10^{-6}$ & 7.2$\times10^{-5}$ \\
Dielectric loss & 2.4$\times10^{-4}$ & 5.7$\times10^{-4}$ \\
Thermal excitation & 1.9$\times10^{-5}$ & 9.3$\times10^{-7}$ \\
Signal separation & 0.15 & 2.7$\times10^{-16}$ \\ \hline
\hline
 & 25 ns flux ramp & 50 ns flux ramp \\ \hline
Non-QNDness & 6.5$\times10^{-4}$ & 4.8$\times10^{-3}$ \\ \hline
\hline
\end{tabular}
\end{table}

\section*{Acknowledgements}
We thank J. Heinsoo and M.F.S. Zwanenburg for insightful discussions.
T.V.S. acknowledges the support of the Engineering and Physical Sciences Research Council (EPSRC) under EP/SO23607/1. C.K.A. additionally acknowledges support from the Dutch Research Council
(NWO).

\section*{Data Availability}
The code used to generate these numerical results is available at  \url{https://github.com/AndersenQubitLab/FluxPulseAssistedFluxonium}.

\section*{Appendix A: Fabrication Variations}

\begin{table*}[t]
\caption{Overview of fluxonium energy and readout parameters used to simulate the performance of flux-pulse-assisted readout with variations from the targeted values of $E_{J}\slash2\pi$ = 4.75 GHz, $E_{L}\slash2\pi$ = 1.5 GHz, and $E_{C}\slash2\pi$ = 1.25, GHz that may result from fabrication. The subscripts SS and RO denote values at the sweet-spot and readout point, respectively.}
\label{tab:readout parameters}
\renewcommand{\arraystretch}{1.5}
\setlength{\tabcolsep}{3.9pt}
\begin{tabular}{lcccccccccc}
\hline\hline
 & $E_{J}\slash2\pi$ & $E_{L}\slash2\pi$ & $E_{C}\slash2\pi$ & $\omega_{q, SS}\slash2\pi$ & $\chi_{SS}\slash2\pi$ & $\Phi_{ext, RO}\slash\Phi_{0}$ & $\omega_{q, RO}\slash2\pi$ & $\chi_{RO}\slash2\pi$ & $g_{20, RO}\slash2\pi$ & $\Delta_{20, RO}\slash2\pi$ \\
Energy variations & GHz & GHz & GHz & GHz & MHz & a.u. & GHz & MHz & MHz & MHz \\ \hline
$E_{J}\slash2\pi$ + 10\% & 5.225 & 1.5 & 1.25 & 0.871 & 0.649 & 0.635 & 4.74 & -7.69 & 20.2 & -85.1 \\
$E_{J}\slash2\pi$ - 10\% & 4.275 & 1.5 & 1.25 & 1.25 & 0.425 & 0.645 & 4.42 & -7.00 & 16.6 & -62.3 \\
$E_{L}\slash2\pi$ + 5\% & 4.75 & 1.575 & 1.25 & 1.12 & 0.493 & 0.633 & 4.49 & -7.86 & 18.2 & -66.4 \\
$E_{L}\slash2\pi$ - 5\% & 4.75 & 1.425 & 1.25 & 0.974 & 0.570 & 0.649 & 4.72 & -7.15 & 18.4 & -76.6 \\
$E_{C}\slash2\pi$ + 5\% & 4.75 & 1.5 & 1.3125 & 1.12 & 0.519 & 0.632 & 4.41 & -7.65 & 18.6 & -74.2 \\
$E_{C}\slash2\pi$ - 5\% & 4.75 & 1.5 & 1.1875 & 0.975 & 0.555 & 0.649 & 4.77 & -7.93 & 17.8 & -61.3 \\ \hline\hline
\end{tabular}
\end{table*}

\begin{figure}
    \includegraphics[scale=1]{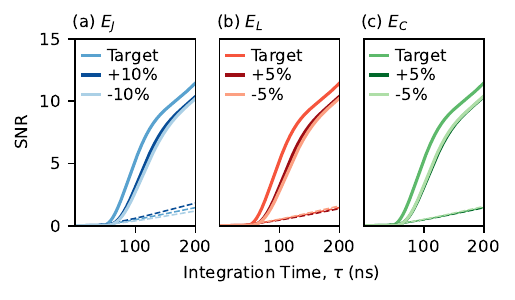}
    \caption{Signal to noise ratio versus integration time for variations in qubit energy parameters from the targeted values of (a) $E_{J}$, (b) $E_{L}$, and (c) $E_{C}$. In changing one energy parameter, the other two are maintained at the targeted values. Solid lines denote the resulting SNR achieved with flux-pulse-assisted readout, utilizing a readout point with comparable parameters (refer to Table ~\ref{tab:readout parameters}) to those used in the main text. Dashed lines show the SNR achieved using static readout at sweet-spot.}
    \label{fig:energy vars}
\end{figure}

Translation of theoretical energy parameters to those of a working device may prove practically challenging. Careful calibration required of fabrication processes, material defects, and device aging may lead to deviations of qubit parameters from targeted values \cite{pop2012fabrication}. In order to look at the robustness of our proposal, we compare the SNR attainable during flux-pulse-assisted readout when one of the fluxonium energy parameters is varied from the value proposed in the main text. We consider an error of $\pm 5\%$ in the case of $E_{L}$ and $E_{C}$, and an error of $\pm 10\%$ in the case of $E_{J}$. We account for greater variation in the Josephson energy as it is more prone to deviations because of the small size of the single junction area that must be precisely targeted \cite{pishchimova2023improving, osman2021simplified, kreikebaum2020improving}. In Fig.~\ref{fig:energy vars}, we compare the achievable SNR using static, sweet-spot (dashed lines) and flux-pulse-assisted (solid lines) readout when each fluxonium energy parameter is independently subjected to a small perturbation.

In each instance, we exploit the dispersive shift feature arising from $\Delta_{20}\rightarrow0$ and ensure the computational subspace is well into the dispersive regime. The exact parameter values are listed in Table~\ref{tab:readout parameters}. The amplitude of the flux pulse is adjusted in each case to achieve comparable dispersive shift, effective coupling, and detuning values at the readout point, and the drive amplitude is updated accordingly. The other relevant readout parameters, including resonator frequency, resonator linewidth, and total coupling, are kept constant.

Although both positive and negative fluctuations in all three fluxonium energy parameters cause slight degradation in the SNR compared to when we have perfect targeting, it is evident that we retain the advantage provided by flux-pulse-assisted readout. The case of $E_{J}\slash2\pi$ = 4.275 GHz yields the largest drop in SNR, yet still reaches an SNR of about 8 after only 155 ns integration time. This value is a substantial improvement compared to the SNR achieved with static readout at sweet-spot.

\section*{Appendix B: Coherent Exchange During Flux Pulse Ramping}
Our flux-pulse-assisted readout scheme with the proposed parameters requires traversing the avoided crossing of the $\ket{3} \leftrightarrow \ket{1}$ qubit transition and readout resonator. We quantify the non-QNDness resulting from the exchange interaction during the 50 ns flux pulse ramp time. We consider the 2-dimensional Hilbert space spanned by $\ket{1, 1}$ and $\ket{3, 0}$. We then construct the time-dependent Jaynes-Cummings Hamiltonian within this subspace
\begin{equation}
    \hat{H}_{\mathrm{JC}, 31}(t) = \frac{\hbar\Delta_{31}(t)}{2}\hat{\sigma}_{z} + \hbar\abs{\alpha_{-}(t)}g_{31}(t)\hat{\sigma}_{x}.
    \label{qnd}
\end{equation}

\begin{figure}
    \includegraphics[scale=0.8]{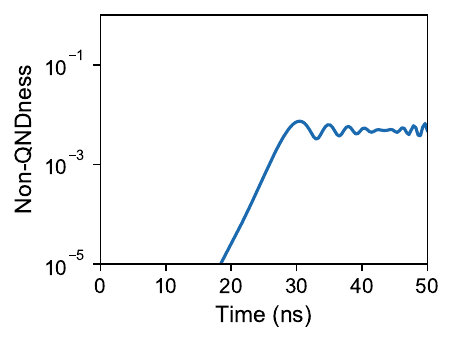}
    \caption{Non-QNDness as a function of time for 50 ns linear ramping of the flux pulse used to tune the qubit. This figure considers the population transferred from the resonator to the third qubit excited state as the avoided crossing around $\Phi_{ext}\slash\Phi_{0}\approx0.575$ is traversed.}
    \label{fig:qndness}
\end{figure}

In this Hamiltonian, $\Delta_{31}$ is the difference between the bare qubit $\ket{3}\leftrightarrow\ket{1}$ transition and bare resonator frequencies and $\alpha_{-}$ is the intracavity resonator field as in Eq.~\eqref{alphasolution}. The time dependence arises as the parameters change while the external flux bias is being linearly ramped from $\Phi_{ext}\slash\Phi_{0}$ = 0.5 to 0.641. Assuming $\ket{1, 1}$ is our initial state, we solve for the population of $\ket{3, 0}$ at the end of the flux ramping via time-evolution generated by Eq.~\eqref{qnd}, see Fig.~\ref{fig:qndness}. We refer to the amount of exchanged state population as non-QNDness.

For a 50 ns rise time, our non-QNDness oscillates around 4.8$\times10^{-3}$ once the $\ket{3}\rightarrow\ket{1}$ avoided crossing has been passed. It is possible to reduce this by shortening the rise time to 25 ns, yielding a non-QNDness that oscillates about 6.5$\times10^{-4}$ post-avoided crossing. However, in doing so we trade off with the adiabaticity of the flux ramping. Finally, pulse shaping of the flux pulse to rapidly cross the avoided crossing may further reduce the non-QNDness.

\section*{Appendix C: Effect of Relaxation Time on Readout Error}
In the main text, we discuss limitations of the flux-pulse-assisted readout scheme that are intrinsic to the readout design. However, relaxation of the qubit inevitably limits the fidelity of fundamental operations, irrespective of readout method. Dielectric loss, which is dictated by structure geometry and material properties, largely limits qubit relaxation times \cite{wang2015surface}. For frequencies above $\sim$ 1 GHz, we can use the dielectric loss model with a frequency independent loss tangent to predict the relaxation rate of the fluxonium qubit according to \cite{sun2023characterization, nguyen2019high, somoroff2021millisecond}
\begin{equation}
    \Gamma_{1} = \frac{\hbar\omega_{q}^{2}}{4E_{C}}\abs{\bra{1}\hat{\phi}\ket{0}}^{2}\mathrm{tan}\,\delta_{C}\mathrm{coth}\left(\frac{\hbar\omega_{q}}{2\mathrm{k_{B}T_{eff}}}\right).
\end{equation}

Assuming a dielectric loss tangent of tan $\delta_{C}$ = 0.8$\times10^{-6}$ and an effective temperature of $\mathrm{T_{eff}}$ = 20 mK, we achieve $T_{1} \approx$ 210 $\mu$s at sweet-spot and $T_{1} \approx$ 77 $\mu$s at the readout point \cite{nguyen2019high, somoroff2021millisecond, zhang2021universal}. We can convert this contribution to readout error in the same manner as Eq.~\eqref{Purcell}, replacing $\Gamma_{\mathrm{P}}$ with $\Gamma_{1}$. After 200 ns integration time, the error contribution to flux-pulse-assisted readout is 5.7$\times10^{-4}$, compared to an error of 2.4$\times10^{-4}$ during readout at sweet-spot. Further reduction in the dielectric loss tangent via attention to materials and fabrication can aid in reducing these errors.

\bibliography{bib.bib}

\end{document}